\documentclass[aps,preprint]{revtex4}%
\usepackage{amsmath}
\usepackage{amsfonts}
\usepackage{amssymb}
\usepackage{graphicx}%
\begin{document}
\title{An interpretation for the entropy of a black hole}
\author{Baocheng Zhang}
\email{zhangbc@wipm.ac.cn}
\affiliation{State Key Laboratory of Magnetic Resonances and Atomic and Molecular Physics,
Wuhan Institute of Physics and Mathematics, Chinese Academy of Sciences, Wuhan
430071, China}
\affiliation{Graduate University of Chinese Academy of Sciences, Beijing 100049, China}
\author{Qing-yu Cai}
\email{qycai@wipm.ac.cn}
\affiliation{State Key Laboratory of Magnetic Resonances and Atomic and Molecular Physics,
Wuhan Institute of Physics and Mathematics, Chinese Academy of Sciences, Wuhan
430071, China}
\author{Ming-sheng Zhan}
\affiliation{State Key Laboratory of Magnetic Resonances and Atomic and Molecular Physics,
Wuhan Institute of Physics and Mathematics, Chinese Academy of Sciences, Wuhan
430071, China}
\affiliation{Center for Cold Atom Physics, Chinese Academy of Sciences, Wuhan 430071, China}
\author{Li You}
\affiliation{Department of Physics, Tsinghua University, Beijing 100084, China}

\begin{abstract}
We investigate the meaning of the entropy carried away by Hawking radiations
from a black hole. We propose that the entropy for a black hole measures the
uncertainty of the information about the black hole forming matter's
precollapsed configurations, self-collapsed configurations, and
inter-collapsed configurations. We find that gravitational wave or
gravitational radiation alone cannot carry all information about the processes
of black hole coalescence and collapse, while the total information locked in
the hole could be carried away completely by Hawking radiation as tunneling.


\end{abstract}

\maketitle

%\tableofcontents

Information is physical and thus it is conserved during any physical process.
Information cannot be simply created out of nothing or disappeared into a sink
hole. The revolution of information science has elevated the above principle
into the status of the fundamental laws of nature \cite{rl91,nc00,sl05}. This
law can be embodied both in classical mechanics and in quantum mechanics.
Classically, a physical state is specified by its distribution function in the
multi-dimensional phase space for all its degrees of freedom. Liouville's
theorem: the conservation of phase space volume, gives rise to the
conservation of entropy or information under Hamiltonian dynamics. In quantum
mechanics, the conservation of information is expressed as the unitarity
evolution for a quantum system, which implies a pure state will evolve into
another pure state, and will never evolve into a mixed state except under
interventions from the external world. In general, the conservation of
information is viewed as the conservation of entropy conservation
quantitatively, \textit{i.e.}, the evolution of an isolated closed system will
not lead to entropy increase or information loss.

The discovery of Hawking radiation from a black hole \cite{swh74,swh75},
however, brings up a serious challenge to the conservation law of information
\cite{swh76}. It was shown that Hawking radiations governed by a thermal
emission causes an increase of entropy after evaporation of a black hole
\cite{whz82}. In other words, information is lost during the process of black
hole evaporation. A revisit of the original treatment for Hawking radiation,
however, revealed that the background geometry was considered as fixed without
enforcing the energy conservation. Including the energy conservation, Parikh
and Wilczek obtained a non-thermal spectrum for Hawking radiation due to
tunneling \cite{pw00}. Along this line, we discovered the existence of
correlations among Hawking radiations by using the standard statistical method
and quantum information theory \cite{zcyz09,zczy09}. By counting the entropy
carried away by emitted particles with themselves, we showed the Hawking
radiation as tunneling is an entropy conserved process. Thus information
remains conserved even during the Hawking radiation from a black hole.
Moreover, along our line in Ref.\cite{zcyz09,zczy09}, the effect of quantum
correction and back reaction for the resolving the information loss paradox
has been investigated \cite{sv10}. It is noted that recently the unitary
character of the black hole evolution is shown from the Schmidt decomposition
viewpoint \cite{bt09}.

In this paper, we investigate the meaning of the entropy carried away by
Hawking radiations from a black hole. Based on the conservation of entropy as
we have discussed before \cite{zcyz09,zczy09}, we also hope to understand the
meaning of the initial black hole entropy. Historically, many interpretations
have been suggested \cite{rmw93,cw94,jj96,sv96,lp97,wjp01,bh07} to explain the
entropy of a black hole entropy, including some novel and profound ideas. Our
explanation comes from the perspective of quantum information and our recent
analysis of Hawking radiation. We interpret entropy as the uncertainty about
the information of the black hole forming matter's precollapsed
configurations, self-collapsed configurations, and inter-collapsed
configurations. The explanation is applied to several circumstances, including
the formation of a black hole, black hole coalescence, and a common matter
dropped into a black hole.

Within the framework of Parikh and Wilczek treatment of Hawking radiation as
tunneling \cite{pw00}, the tunneling probability is found to be nonthermal and
given by
\begin{align}
\Gamma(M;E) \sim\exp\left[  -8\pi E\left(  M-\frac{E}{2}\right)  \right]  .
\end{align}
Quite straightforwardly, the exponential part can be considered as the entropy
change of a black hole, $\Delta\mathcal{S}=-8\pi E\left(  M-{E}/{2}\right)  $.
The negative sign represents the decrease of the black hole entropy associated
with each emission. This implies information is carried away by Hawking
radiation because a reduced entropy implies a reduced uncertainty or the
gaining of information. According to information theory, the entropy carried
away by emitted particles is defined by $S(E)=-\ln\Gamma(M;E)=+8\pi
E(M-{E}/{2})$, where the positive sign represents an increase of the entropy
for the environment surrounding a black hole. We adopt a notation using
$\mathcal{S}_{m}$ and $S(E)$ to denote the entropies respectively for a black
hole of mass $m$ and for a radiation of energy $E$. Thus for the complete
system of a black hole plus its Hawking radiations, entropy is not changed
during the whole process, although information is indeed carried to the
outside a black hole.

At first, we will explain why the tunneling particle could take the entropy by
itself in Hawking radiation as tunneling. The most important reason is that
the emission process is probabilistic, not deterministic. For each tunneling
emission from a black hole with the mass $M$, we only know a radiation may
occur with a probability $\Gamma(M;E)$, nothing else. In other words, the
uncertainty of the event (for a radiation with energy E) or the potential
information we can gain from the event is $S(E)=-\ln\Gamma(M;E)$.

Proceeding with an explicit presentation, we rewrite the entropy \cite{zcyz09}
carried away by the emitted particle with an energy $E$ as%
\begin{align}
S(E)=8\pi E\left(  M-\frac{E}{2}\right)  =8\pi E(M-E)+4\pi E^{2}, \label{pen}%
\end{align}
which depends not only on the energy of the emitted particle, but also on the
mass of the black hole. After the emission of a particle with an energy $E$,
the whole system consists of a new black hole with mass $M-E$ and an emission
with energy $E$. The first term in the Eq. (\ref{pen}) is identical in form to
the correlation between the black hole and the particle. The second term in
Eq. (\ref{pen}) is the familiar entropy of a black hole with mass or energy
$E$. This understanding can be further clarified by considering the coalescing
of two black holes.

We next consider two Schwarzschild black holes with respective mass $M$ and
$m$. Their respective entropies are $4\pi M^{2}$ and $4\pi m^{2}$. Assuming
they are a large distance apart initially and are being held stationary, both
the total kinetic energy and momentum can be taken as zero. Due to
gravitational attractive interaction, the two black holes will approach each
other with ever increasing velocity until they experience a head-on collision.
If the collision is elastic, the total kinetic energy and momentum will be
conserved, the two black holes will not coalesce into one larger black hole.
The above picture is thus not in conflict with any conservation laws of
fundamental physics, such as energy conservation, momentum conservation, and
entropy conservation. But it is not a realistic scenario as two colliding
black holes will form a larger black hole. While energy conservation and
momentum conservation is strictly held, entropy is not conserved when two
black holes coalesce into one as we can easily check by writing down the
entropy of the resulting black hole as
\begin{align}
\mathcal{S}_{M+m}=4\pi(M+m)^{2}=4\pi M^{2}+4\pi m^{2}+8\pi Mm.
\end{align}
The extra (third) term $8\pi Mm$ measures some kind of correlation generated
by gravitational interactions. Before the two coalesce, this correlation
constitutes of actual information describing dynamics due to gravitational
force. It can be gained by an exterior observer, so the entropy for the whole
system will not change. After the new black hole forms, the correlation is
covered by the resulting event horizon, the exterior observer will not be able
to obtain this information about correlation, so that entropy increases, or
the uncertainty for the new system (the new black hole) increases.

On the other hand, when the two black holes collide and coalesce into one,
gravitational waves are emitted. Is it possible that the gravitational
radiations actually carry away the amount of information corresponding to the
increased entropy? If one takes an affirmative attitude towards this question,
the entropy carried away by gravitational radiations has to be at least of the
following magnitude
\begin{align}
S(m^{\prime})  &  = \mathcal{S}_{M}+\mathcal{S}_{m}-\mathcal{S}_{M+m-m^{\prime
}}\nonumber\\
&  = 8\pi(M+m)m^{\prime}-4\pi{m^{\prime}}^{2}-8\pi Mm,
\end{align}
where $m^{\prime}$ is the energy of gravitational wave radiation.

According to the classical area theorem \cite{bch73}, when two black holes
coalesce, the area of the final event horizon is greater than the sum of the
areas of the initial horizon. Thus, the entropy for the new black hole will be
greater that the sum of the entropies for the two initial black holes, because
the entropy for a black hole is proportional to its area of the event horizon.
This gives the following inequality $\mathcal{S}_{M+m-m^{^{\prime}}%
}>\mathcal{S}_{M}+\mathcal{S}_{m}$, from which, we find $S(m^{\prime})<0$.
Gravitational radiation thus cannot carry away all the increased entropy.
There must exist other correlations covered by the event horizon of the
coalesced black hole and inaccessible to the exterior observers. This leads to
the increase of entropy. In other words, gravitational wave radiations alone
cannot carry all the information about the gravitational interactions during
the collapse.

More generally, we consider a common matter of mass $m$ falling a black hole.
The entropy for the resulting black hole can be expressed as $\mathcal{S}%
_{M+m}=4\pi\left(  M+m\right)  ^{2}=4\pi M^{2}+4\pi m^{2}+8\pi Mm$ due to
conservation of energy. If the initial entropy of the fallen matter is
$\mathcal{S}^{(0)}$, the net entropy increase is
\begin{align}
\Delta\mathcal{S}=4\pi m^{2}+8\pi Mm-\mathcal{S}^{(0)}. \label{mec}%
\end{align}
Without the detailed knowledge for the microstate of the fallen mass, it is
impossible to estimate its entropy. However, the expression for the entropy
change (\ref{mec}), suggests the description of the process for a matter
falling into a black hole can be generally separated into two stages: 1), the
fallen matter becomes a black hole in a self-collapsed process. This is
analogous to how a common mass $m$ would collapse into a black hole. The
entropy increase reveals the inaccessible information about the collapse. In
quantitative terms, this increased entropy is given by $4\pi m^{2}%
-\mathcal{S}^{(0)}$; 2), the initial black hole and the black hole of the
fallen mass coalesce into a new black hole in an inter-collapsed process. The
process is always accompanied by the emissions of gravitational waves. Hawking
once obtained an upper bound of 29\% for the total energy of gravitational
waves emitted when one collapsed object captures another \cite{swh71}.
Recently this upper bound has reduced appreciably based on numerical
simulations of the Einstein's equation \cite{scp08}. The previous
considerations as discussed above show that gravitational wave radiations
cannot carry away all the increased entropy. Therefore this inter-collapsed
process leads to an increase of entropy due to the inaccessible information
about the correlations during the coalescence and collapse. This increased
quantity is exactly given by $8\pi Mm$.

The above general case for a common matter falling into a black hole also
applies to the case of a common matter collapsing into one black hole. We can
simply view this process as first from individual parts of the common matter
forming individual baby black holes; the baby black holes interact and merge
with each other and finally coalesce and form a new larger black hole.

In the following we proceed to investigate the meaning of the entropy
(\ref{pen}). Before the formation of a black hole, we denote the entropy for a
particle (mass) with energy $E$ by $\mathcal{S}^{(0)}$, the entropy
(\ref{pen}) is then conveniently reexpressed as
\begin{align}
S(E)=8\pi E(M-E)+(4\pi E^{2}-\mathcal{S}^{(0)})+\mathcal{S}^{(0)}.
\end{align}
Thus this entropy, which measures the information carried away by the
tunneling particle, measures respectively its inter-collapsed configurations,
self-collapsed configurations, and the precollapsed configurations. In the
radiation process, in addition to the information or entropy $\mathcal{S}%
^{(0)}$, inherent to the radiating particle, the correlation between the
radiation and the remaining black hole $8\pi E\left(  M-E\right)  $, generated
from the inter-collapsed process, and entropy of the remaining black hole
$\left(  4\pi E^{2}-\mathcal{S}^{(0)}\right)  $, generated by the
self-collapsed process, are carried away as well.

The exterior correlations include correlations of all emitted particles with
each other, which can be shown for any queue of Hawking radiations as
sequential tunneling. For this purpose, we consider a queue of emissions
ordered according to $E_{1}$, $E_{2}$, $\cdots$, $E_{n-1}$. The entropy of the
first emission with an energy $E_{1}$ is
\[
S(E_{1})=8\pi E_{1}(M-E_{1})+(4\pi E_{1}^{2}-\mathcal{S}_{1}^{(0)}%
)+\mathcal{S}_{1}^{(0)},
\]
where the term $8\pi E_{1}(M-E_{1})$ includes all the correlations between the
particle with energy $E_{1}$ and all other particles with energies $E_{2}$,
$\cdots$, $E_{n-1}$, $E_{n}$. Given the first emission with an energy $E_{1}$,
the entropy of the second emission with an energy $E_{2}$ is
\begin{align}
S(E_{2}|E_{1})  &  =8\pi E_{2}(M-E_{1}-E_{2})\nonumber\\
&  +(4\pi E_{2}^{2}-\mathcal{S}_{2}^{(0)})+\mathcal{S}_{2}^{(0)}. \label{2es}%
\end{align}
As before, this entropy is partitioned into three terms with $\mathcal{S}%
_{2}^{(0)}$ referring to the precollapsed configurations, $(4\pi E_{2}%
^{2}-\mathcal{S}_{2}^{(0)})$ about self-collapsed configuration, and the
correlation, or partial information $8\pi E_{2}(M-E_{1}-E_{2})$ about
inter-collapsed configuration. The information about the correlations between
the first emission $E_{1}$ and the second emission $E_{2}$ is already carried
out by the first emission. Therefore for the second emission (\ref{2es}), its
correlation with the first $8\pi E_{1}E_{2}$ must be subtracted. Analogously,
for the third emission with energy $E_{3}$,
\begin{align}
S(E_{3}|E_{1},E_{2})  &  =8\pi E_{3}(M-E_{1}-E_{2}-E_{3})\nonumber\\
&  +(4\pi E_{3}^{2}-\mathcal{S}_{3}^{(0)})+\mathcal{S}_{3}^{(0)}.
\end{align}

It is easy to check that correlations between the third emission $E_{3}$ and
first one $E_{1}$, and between the third emission $E_{3}$ and the second
emission $E_{2}$ are already subtracted. We summed together, $S(E_{1}%
)+S(E_{2}|E_{1})+S(E_{3}|E_{1},E_{2})$ contains no redundant information or
entropy. Thus, this sum of entropies is equivalent to the reduced entropy for
the black hole, or
\begin{align}
&  S(E_{1})+S(E_{2}|E_{1})+S(E_{3}|E_{1},E_{2})\nonumber\\
&  =4\pi(M-E_{1}-E_{2}-E_{3})^{2}-4\pi M^{2}\nonumber\\
&  =\Delta S_{\mathrm{BH}}.
\end{align}

This step by step construction shows that the Hawking radiations carry with
themselves information, in fact, all information, because no information loss,
or entropy increase is found. Additionally, our analysis above provides a
self-consistent interpretation for the entropy of a black hole according to
the information of entropies taken out by the Hawking radiations. The entropy
for a black hole merely implies that for an exterior observer, there exists
uncertainties for the information about precollapsed configurations,
self-collapsed configurations, and inter-collapsed configurations. When a
black hole radiates, all these associated information are leaked out through
the particles and the correlations between particles.

Repeating the above process of step by step analysis of each Hawking emissions
until the black hole is completely exhausted, the entropy or information
conservation is found to be preserved at all times. For a Schwarzschild black
hole, however, it is difficult to describe the final emission $E_{n}$, whose
entropy is
\[
S(E_{n}|E_{1},E_{2}\cdots,E_{n-1})=4\pi E_{n}^{2},
\]
which is precisely the same as for a black hole with mass or energy $E_{n}$.
This shows the final emission is really equivalent to emit itself, because
when the black hole is about to vanish due to evaporation, the temperature
becomes very high enough to emit the particle with any mass or energy.
Moreover, noted that the final black hole could be regarded as a fundamental
particle \cite{gh85,yh09}, which is stable and emit no radiation. In Ref.
\cite{zczy09}, when quantum gravity effects \cite{amv05} are considered, the
black hole will evolve into a remnant and the problem of an infinite
temperature is voided. Whichever case happens, it seems that our conclusion of
entropy conservation in the Hawking radiation process is unaffected.

In conclusion, based on carefully analyzing the entropies carried away by
tunneling particles, we find the black hole entropy contains three parts:
respectively associated with the information for precollapsed configurations,
self-collapsed configurations, and inter-collapsed configurations. All
information are covered by the event horizon of a black hole and are
inaccessible to exterior observers. When a black hole emits, all such
information are taken out of the black hole by the radiations, and this
implies that the black hole evaporation is a unitary process.

When two black holes coalesce to form a new black hole, the gravitational
waves emitted during the process are found to be incapable of carrying away
information associated with the increased entropy. This implies one cannot
obtain all the information about the collapsing process by gravitational
radiations or gravitational waves. It sounds disappointing. However, our work
suggest that Hawking radiations, on the other hand, contain all information
about the gravitational collapse.

A final comment concerns the following question: why the entropy of an
ordinary matter, which could essentially take any value, changes into a fixed
value after fallen/changed into a black hole of the equal energy? As pointed
out by some physicists \cite{jdb81,ls95,rb99,fmw00}, black holes have the
maximum possible entropy of any object of equal size and this makes them
likely end points of all entropy-increasing processes. We don't attempt to
prove it or provide an our answer in this paper. Instead, we simply provide an
explanation for the increased entropy, which we feel suggests that information
about gravitational interaction or gravitational spacetime is closed inside
the event horizon. This shows that the guess of the maximum entropy is at
least not in conflict with information conservation. Of course, a clearer
explanation about black hole entropy need a better description for the state
of the inner black hole. Although string theory and many other quantum gravity
theories can give some such descriptions, the price paid includes additional
elements not completely falsifiable at the present stage.

This work is supported by National Basic Research Program of China (NBRPC)
under Grant No. 2006CB921203.

\end{document}